\documentstyle[aps,preprint]{revtex}
\tightenlines

\begin{document}
\draft
\title{Shapiro steps in a superconducting film with an antidot lattice}
\author{L. Van Look, E. Rosseel, M. J. Van Bael, K. Temst, V.~V.~Moshchalkov and Y.~Bruynseraede}
\address{Laboratorium voor Vaste-Stoffysica en Magnetisme,\\
Katholieke Universiteit Leuven\\
Celestijnenlaan 200D, B-3001 Leuven, Belgium}
\date{\today}
\maketitle

\begin{abstract}

Shapiro voltage steps at voltages $V_n=nV_0$ ($n$ integer) have been observed in the voltage-current  characteristics of a superconducting film with a square lattice of perforating microholes (antidots) in the presence of radiofrequent radiation. These equidistant steps appear at the second matching field $H_2$ when the flow of the interstitial vortex lattice in the periodic potential created by the antidots and the vortices trapped by them, is in phase with the applied rf frequency. Therefore, the observation of Shapiro steps clearly reveals the presence of mobile intersitial vortices in superconducting films with regular pinning arrays. The interstitial vortices, moved by the driving current, {\it coexist with immobile vortices} strongly pinned at the antidots.
\end{abstract}



Conventional Josephson junctions (JJ's), irradiated with a radiofrequent signal, show Shapiro steps in the voltage current ($V(I)$) characteristics\cite{Shapiro}, which are equidistant in voltage and show a constant ratio $\nu/V_1=h/2e$, with $V_1$ the height of the first step and $\nu$ the irradiation frequency. In the framework of the resistively shunted junction (RSJ) model, the behavior in time of $\Delta\Phi$, the phase difference between the two electrodes of a JJ, can be described by the damped pendulum equation\cite{Waldram}. This equation of motion is completely analoguous to the one of a driven mass on a tilted washboard surface. A horizontal driving force on this mass, corresponding to the current through the JJ, is represented by the tilt of the washboard. From this analogy, the origin of the voltage steps can be interpreted as follows: the dc driving force on the mass determines the average tilt of the washboard while the ac force makes the tilt oscillate around this average position with the applied ac frequency. When the period (or a multiple of the period) of the hopping of the mass over the barriers of the washboard coincides with the period of the applied ac force, an interference effect occurs, resulting in steps in the $V(I)$-characteristic. 

	This appearance of Shapiro steps is therefore not exclusively seen in the $V(I)$-characteristics of rf-irradiated JJ's but is expected in every system where an object, moving in a periodic potential, is driven by a superimposed dc and ac force. 	An interesting example of such a system is the dc flux transformer\cite{Sherril,Gilabert}.

	Another system in which Shapiro steps can be observed, is a superconducting film with a laterally modulated thickness. Here the Abrikosov vortex lattice moves coherently in the {\it one-dimensional periodic potential} created by the thickness modulation. Martinoli et al.\cite{Martinoli} found pronounced equidistant steps in the $V(I)$-characteristics of these thickness-modulated films at several matching fields and rf frequencies.	

	In this paper, we focus on the observation of Shapiro steps in a superconducting film with a {\it two-dimensional periodic potential} created by a lattice of antidots, i.e. sub-micron holes (see Fig.~\ref{fig:afm}a). Chosing the appropriate temperature and magnetic field, this system can be tuned so, that every antidot is occupied by a single vortex and an interstitial vortex lattice is formed at the centres of the cells, caged by the surrounding occupied antidots (see Fig.~\ref{fig:afm}b)\cite{Rosseel2}. This weakly pinned interstitial lattice is easily moved through the potential that is created by the strongly pinned vortices at the antidots. We present $V(I)$ results obtained at the second matching field $\mu_0H_2$ on a superconducting Pb film containing a square array of antidots. We show clear evidence for the existence of Shapiro steps in these films. Moreover, the height of the appearing voltage steps proves that only the interstitial vortex lattice is moving, in agreement with recent Lorentz-microscopy experiments\cite{Harada}.


	The studied samples are 50 nm thick Pb films with a square antidot lattice (antidot size $a^2$ = 0.4 x 0.4 $\mu$m$^2$ and period $d$ = 2 $\mu$m). Fig.~\ref{fig:afm}a shows an atomic force micrograph of the sample. The films are electron beam evaporated in a MBE apparatus onto liquid nitrogen cooled SiO$_2$ substrates with a predefined resist-dot pattern and are covered with a protective Ge layer of 20 nm. The films have a strip geometry of 0.3 x 3 mm$^2$ with two current contacts and three equally spaced voltage contacts on each side of the strip. The part of the strip which lies between the used voltage contacts is 2 mm long and contains 1000 rows of antidots. From the $T_c(H)$ fase boundary of a reference plain film, coevaporated with the antidot lattice, the superconducting coherence length was determined to be $\xi$(0)=38 nm \cite{Rosseel}. The samples are measured in a $^3$He cryostat using a DC current source (Keithley 238) and a nano-voltmeter (Keithley 182). The rf signal was generated by a 9 kHz-2 GHz signal generator (Rohde $\&$ Schwarz SMY 02). Fig.~\ref{fig:setup}a shows a schematic drawing of the measuring setup. The rf signal was superimposed on the dc current through two 100 nF capacitors.


	Due to the applied rf current and the associated Lorentz force, the pinning potential is tilted periodically, and when the resulting flow of the vortex lattice is in phase with the rf modulation of the pinning potential, steps occur in the $V(I)$ characteristic at well defined voltages. For a square lattice of moving vortices with $k$ moving vortices per unit cell of the antidot array, these voltages are given by\cite{Martinoli} :

\begin{equation}
V_n=Nk\frac{h}{2e}\frac{\langle v \rangle}{d}=n (Nk \frac{h}{2e} \nu)=n(Nk\Phi_0 \nu)\equiv n V_0  \label{Eq:Vn}
\end{equation}

where $n$ is an integer, $N$ is the number of antidot rows between the voltage contacts ($N$=1000 for this sample), $\langle v \rangle$ is the average velocity of the coherent motion of the interstitial vortex lattice, $\nu$ is the frequency of the applied rf signal and $d$ is the period of the potential created by the singly occupied antidots. The third equality can be obtained from the second using the fact that, when the rf signal and the vortex flow are in phase, $\langle v \rangle /d$=$n\nu$. At $n$=1 the vortices propagate over one lattice period $d$ during one rf cycle, for $n$=2 they move over two periods during one rf cycle, etc. .  By comparing the observed step height $V_0$ with the one expected from Eq.~(\ref{Eq:Vn}), the number $k$ of moving vortices per unit cell can be determined. This technique can therefore be used to detect the presence and perhaps even the amount of interstitial vortices in superconducting films with an artificial periodic pinning array. 

	In Fig.~\ref{fig:vicurve}  we show a $V(I)$ curve at $T$ = 7.151 K = 0.995 $T_c$  obtained for a sinusoidally modulated current : $I$ = $I_{dc} + I_{rf} \sin(2\pi \nu t)$ where $I_{rf}\simeq I_c$, the critical current of the sample, and $\nu$=40 MHz. The field was fixed at the second matching field $\mu_0H_2$=1.03 mT, determined from previous critical current measurements \cite{Rosseel}. The $V(I)$ curve is nearly a straight line through the origin, with smooth periodic steps superimposed on it. By plotting the derivative $\delta I/ \delta V$ versus $V$, the steps appear as peaks and their position is well defined. In the curve shown in Fig.~\ref{fig:vicurve} the steps have a voltage separation $V_0$=$N k \nu \frac{h}{2e}$ of 81.3 $\mu$V, which is within 2 \% of the value 82.8 $\mu$V calculated from Eq.~(\ref{Eq:Vn}), using $k$=1. This leads to an average flow velocity of interstitial vortices $\langle v \rangle$ of 80 m/s and a traveling time between entrance and exit of $\approx $ 3.75 $\mu$s. 
	
	We have repeated the same experiment for different frequencies down to 25 MHz while keeping the other parameters (temperature, magnetic field, amplitude $I_{rf}$ of the irradiation) fixed. According to Eq.~(\ref{Eq:Vn}), $V_0$ is a linear function of the frequency $\nu$ with a slope of $k$ $\times N \Phi_0$, which is equal to $k$ $\times$ 2.07 $\mu$V/MHz for this sample. The measurements reveal a slope of 2 $\mu$V/MHz (see the upper-left inset of Fig.~\ref{fig:vicurve}), indicating that there is only one moving vortex per unit cell of the antidot array. This means that indeed two vortices per unit cell are present, but that only the interstitial vortices are moving coherently, while the other half of the vortices is pinned to the antidots or creeping incoherently (see the schematic plot in Fig.~\ref{fig:setup}b). 

Indeed, the saturation number \cite{Buzdin} $n_s \approx \frac{r}{2 \xi (T)} = \frac{r}{2 \xi(0)} \sqrt{1- \frac{T}{T_c}}$ = 0.18 indicates that only one vortex can be pinned per antidot, forcing the excess vortices to occupy the much weaker pinning sites at interstices \cite{Baert,Mosh,Mosh2}. Recent Lorentz-microscopy experiments\cite{Harada} and numerical simulations\cite{Reichhardt} have shown that, for a range of driving forces, the motion of interstitial vortices is confined to 1D channels between the adjacent antidot rows. When all interstitial positions are filled, as in our experiments, it is to be expected that the repulsive interaction within the 1D vortex channel and the shear between neighboring rows will lead to a coherent motion of all interstitial vortices, as we observe in our experiments.


In summary, we have shown that Shapiro steps at voltages $V_n=nV_0$ are present in the $V(I)$ curves of a superconducting film with a square antidot lattice when a rf current is superposed on the dc transport current. These voltage steps are equidistant and the ratio $V_0/\nu$ depends on the number of antidot rows enclosed between the voltage contacts and the number of moving vortices per antidot lattice unit cell. From the height of the steps, we were able to  prove that only the interstitial vortices are moving, while the vortices occupying the antidots are strongly pinned or creeping incoherently. Therefore, the presence of Shapiro steps should be attributed to the coherent motion of interstitial vortices in superconducting films with a regular array of strong pinning sites.

\acknowledgments
The authors would like to thank A. Gilabert for useful discussions. This work has been supported by the Belgian Interuniversity Attraction Poles (IUAP) and the Flemish Concerted Action (GOA), the TOURNESOL and ESF 'VORTEX'  programs. M. J. Van Bael is a Postdoctoral Research Fellow of the Fund for Scientific Research (FWO-Vlaanderen).


\begin{figure}[tbp]
\caption{(a) Atomic force micrograph of the Pb(50 nm) film containing a square array of holes (antidots). The antidot size $a$ is 0.4 $\mu$m and the lattice period $d$ is 2 $\mu$m. (b) Schematic representation of the free energy along a diagonal cut of the square antidot lattice. The black and grey arrows represent the vortices strongly pinned at the antidots and the weakly pinned interstitial vortices, respectively. }
\label{fig:afm}
\end{figure}

\begin{figure}[tbp]
\caption{(a) Schematic drawing of the measuring setup. The rf current is superposed on the dc current through the sample via two 100 nF capacitors. (b) Scheme of the configuration of the vortices at the second matching field $\mu_0H_2$. The open squares indicate the position of the antidots, the filled circles represent vortices. Every antidot is singly occupied by a vortex, the interstitial vortex lattice is dragged by the transport current in the direction of the arrows.}
\label{fig:setup}
\end{figure}

\begin{figure}[tbp]
\caption{$V(I)$-curve of a Pb(50 nm) film with a square antidot lattice at $T$ = 7.151 K = 0.995~$T_c$ and $\mu_0H$=$\mu_0H_2$=1.03 mT. The inset shown in the right corner shows the derivative $\delta I/\delta V$  vs. $V$. The horizontal arrows mark the voltage steps expected at multiples of $V_0$=81.3 $\mu $V (Eq.~(\ref{Eq:Vn})). The inset in the upper-left corner shows the linear dependence of $V_0$ upon frequency. A slope of $V_0/\nu$=2$\mu$V/MHz is obtained, proving that only the interstitial vortex lattice is moving.}
\label{fig:vicurve}
\end{figure}


\begin{references}


\bibitem{Shapiro} S. Shapiro, Phys. Rev. Lett. {\bf 11}, 80 (1963).

\bibitem{Waldram} J. R. Waldram, A. B. Pippard, and J. Clarke, Phil. Trans. Roy. Soc. Lond. A. {\bf 268}, 265 (1970).

\bibitem{Sherril} M.D. Sherril and W. A. Lindstrom, Phys. Rev. B {\bf 11}, 1125 (1975).

\bibitem{Gilabert} A. Gilabert, I. K. Schuller, V. V. Moshchalkov, and Y. Bruynseraede, Appl. Phys. Lett. {\bf 64}, 2885 (1994).

\bibitem{Martinoli} P. Martinoli, O. Daldini, C. Leemann, and E. Stocker, Solid State Commun. {\bf 17}, 205 (1975).

\bibitem{Rosseel2} E. Rosseel, M. J. Van Bael, M. Baert, R. Jonckheere, V. V. Moshchalkov, and Y. Bruynseraede, Phys. Rev. B {\bf 53}, R2983 (1996).

\bibitem{Harada} K. Harada, O. Kamimura, H. Kasai, T. Matsuda, A. Tonomura, and V.V. Moshchalkov, Science {\bf 274}, 1167 (1996).

\bibitem{Rosseel} E. Rosseel, Ph.D. Thesis {\it Critical parameters of superconductors with an antidot lattice}, Katholieke Universiteit Leuven, Leuven (1998).

\bibitem{Buzdin} A. Buzdin and D. Feinberg, Physica C {\bf 256}, 303 (1996).

\bibitem{Baert} M. Baert, V. V. Metlushko, R. Jonckheere, V. V. Moshchalkov, and Y. Bruynseraede, Phys. Rev. Lett. {\bf 74}, 3269 (1995).

\bibitem{Mosh} V. V. Moshchalkov, M. Baert, V. V. Metlushko, E. Rosseel, M. J. Van Bael, K. Temst, R. Jonckheere, and Y. Bruynseraede, Phys. Rev. B {\bf 54}, 7385 (1996).

\bibitem{Mosh2} V. V. Moshchalkov,  M. Baert, V. V. Metlushko, E. Rosseel, M. J. Van Bael, K. Temst, Y. Bruynseraede, and R. Jonckheere, Phys. Rev. B {\bf 57}, 3615 (1998).

\bibitem{Reichhardt} C. Reichhardt, C. J. Olson, and F. Nori, Phys. Rev. Lett. {\bf 78}, 2648 (1997).


\end{references}
\end{document}